\documentclass{article}
\usepackage{spconf,amsmath,graphicx,hyperref,booktabs,multirow}
\usepackage[caption=false]{subfig}

\let\oldthebibliography\thebibliography
\renewcommand{\thebibliography}[1]{%
  \oldthebibliography{#1}%
  \setlength{\itemsep}{0pt}%
  \setlength{\parsep}{0pt}%
  \renewcommand{\baselinestretch}{0.9}\selectfont
}

\title{Beyond Intelligibility: Fine-Grained Perceptual Assessment of Dysarthric Speech with Auditory Large Language Models}
\name{Haotian Wei$^{1\star}$, Can Long$^{1\star}$, Wen Xian Jim$^{1\star}$\thanks{$^{\star}$Equal contribution was made between the three authors.}, Guangzhi Sun$^{2}$, Zengrui Jin$^{1}$, Chao Zhang$^{1}$}
\address{
$^{1}$Tsinghua University,
$^{2}$University of Cambridge\\
\texttt{\{whd24,longc24,rwx23\}@mails.tsinghua.edu.cn;}\\
\texttt{gs534@cam.ac.uk; \{zrjin,cz277\}@tsinghua.edu.cn}
}
\begin{document}
\ninept

\maketitle
\begin{abstract}
Accurate assessment of dysarthric speech is vital for clinical diagnosis and rehabilitation, yet existing automated methods rely heavily on isolated, single-task models or generic speech quality metrics that fail to disentangle heterogeneous neuromotor deficits. 
This work presents the first systematic investigation into adapting auditory large language models (ALLMs) for multi-dimensional dysarthric speech assessment under a unified, prompt-conditioned framework. Four representative ALLMs are benchmarked across five clinical perceptual dimensions from the Speech Accessibility Project (SAP): naturalness, monopitch, intelligibility, imprecise consonants, and inappropriate silences. 
To bridge generic perceptual quality and clinical evaluation, a two-stage training paradigm is explored, leveraging out-of-domain pre-training followed by in-domain clinical adaptation. 
Experimental results demonstrate that leading ALLMs reliably capture fine-grained pathological variations, exhibiting distinct and complementary strengths across prosodic and articulatory attributes. 
These findings establish ALLMs as a scalable foundation for clinical speech evaluation and provide practical insights into model selection and multi-task transfer potential.
\end{abstract}
\begin{keywords}
Dysarthric speech assessment, auditory large language models, speech quality
\end{keywords}
\section{Introduction}
\label{sec:intro}

Dysarthria is a neuromotor speech impairment caused by neurological injury or conditions such as cerebral palsy, amyotrophic lateral sclerosis, Parkinson's disease, or stroke \cite{rudzicz2011tasl,rudzicz2009icassp,bhat2025overview}. These motor deficits give rise to distinct perceptual features, among which diminished speech intelligibility and compromised naturalness exert the most severe impact on daily communication \cite{jia2026augmenting}.

Accurate assessment of dysarthric speech remains an open challenge with profound clinical and technological implications, despite recent progress in speech processing \cite{hadian2018end,gulati20_interspeech,yao2024zipformer,yao2025crctc} and dysarthric speech recognition \cite{jin2023personalized,wang23y_interspeech,wang2024enhancing}. Such assessment serves as a sensitive biomarker for the early detection and tracking of neurological disease progression \cite{10095664}, offers objective metrics to guide speech therapy and evaluate rehabilitation outcomes \cite{bhat2017icassp,merler2025clinical}, and plays a critical role in optimizing downstream tasks, including dysarthric speech recognition and disordered speech reconstruction \cite{geng23b_interspeech,jeon25_interspeech,10445949,wang2024unit}.


Despite its clinical significance, evaluating dysarthric speech still relies overwhelmingly on subjective perceptual ratings by certified speech-language pathologists \cite{rudzicz2012torgo,kim08c_interspeech,stipancic2023perceptual}, a process that is labor-intensive, scarce, and inherently difficult to scale.
To alleviate these bottlenecks, automated assessment methods have emerged as an active area of research. However, existing automated systems predominantly rely on isolated, single-task models designed strictly for either intelligibility scoring or overall severity classification \cite{bhat2017icassp,javanmardi2023wav2vec,yeo2023icassp,kadirvelu2025speaker}. 
While recent efforts have drawn upon general speech-quality modeling (e.g., mean opinion score, MOS) to provide complementary supervision for clinical ratings \cite{jia2026augmenting,lo19_interspeech,huang22f_interspeech,mittag21_interspeech,saeki22c_interspeech,wang-etal-2025-qualispeech}, general quality metrics and clinical dysarthria evaluations capture fundamentally distinct constructs. 
General MOS protocols primarily summarize listener impressions under generic degradation, whereas clinical dysarthria assessment strictly demands diagnostic criteria grounded in neuromotor speech production deficits and their communicative impact \cite{green04_interspeech,rudzicz2011tasl,rudzicz2009icassp,bhat2025overview}. 
Consequently, existing paradigms lack a unified mechanism capable of disentangling and quantifying these heterogeneous perceptual dimensions.

Modern Auditory Large Language Models (ALLMs) present a compelling opportunity to bridge this gap. 
By shifting machine audition from narrow pattern matching toward instruction-conditioned general understanding \cite{wang2026towards,tang2024salmonn,chu2024qwen2audio,yang2024airbench,sakshi2025mmau}, ALLMs offer a flexibility to unify multiple assessment targets under a single prompt driven backbone. 
Moreover, their recent adaptation to subjective speech quality modeling has demonstrated substantial perceptual sensitivity \cite{wang2025enabling,chen2025audio}. 
Nevertheless, pathological speech poses an uncharted challenge: its clinical information is governed not by semantic transcripts, but by motor-speech production nuances spanning articulation, phonation, prosody and rate \cite{rudzicz2011tasl,rudzicz2009icassp}. 
It remains an open question whether the general perceptual capabilities of ALLMs can reliably disentangle and quantify fine-grained clinical dimensions of dysarthric speech.

To address this gap, this work investigates whether ALLMs can serve as a unified, prompt-conditioned framework for multi-dimensional dysarthric speech assessment. Our contributions are threefold:
\textbf{ALLM based dysarthria assessment:} To our knowledge, this is the first study to systematically adapt auditory LLMs to multi-dimensional, utterance-level assessment of pathological speech.
\textbf{Unified multi-task formulation:} results demonstrate that complementary perceptual dimensions, ranging from holistic intelligibility to fine-grained phonetic and prosodic qualities, can be jointly modeled within a single instruction-conditioned architecture.
\textbf{Cross-model and cross-dimension investigation:} Extensive benchmarking was conducted across multiple models, parameter scales and clinical dimensions to reveal the capability of auditory LLMs in clinical speech evaluation.

\begin{figure*}[!h]
    \centering
    \subfloat[Naturalness]{\includegraphics[width=0.2\linewidth]{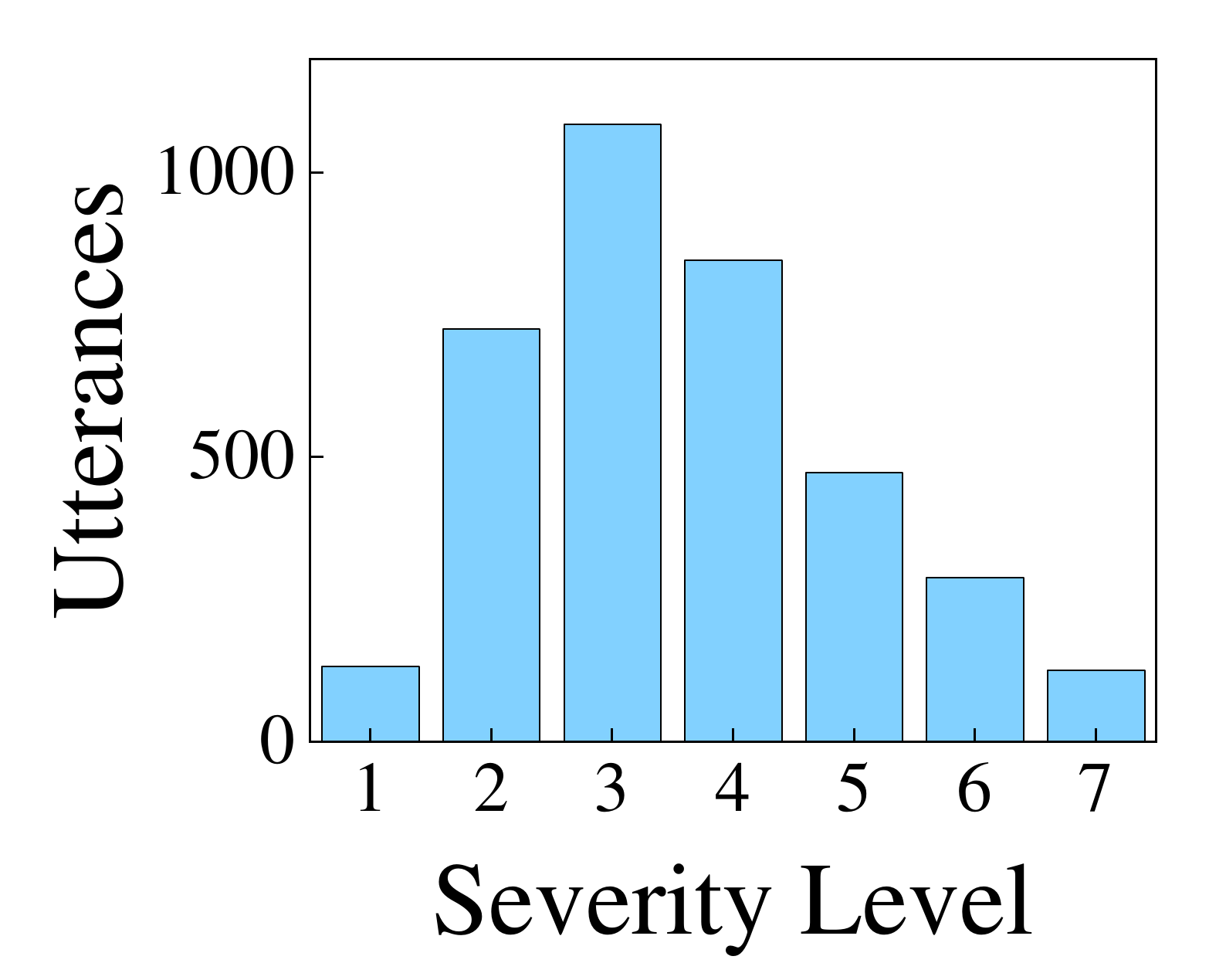}}%
    \hfill
    \subfloat[Monopitch]{\includegraphics[width=0.2\linewidth]{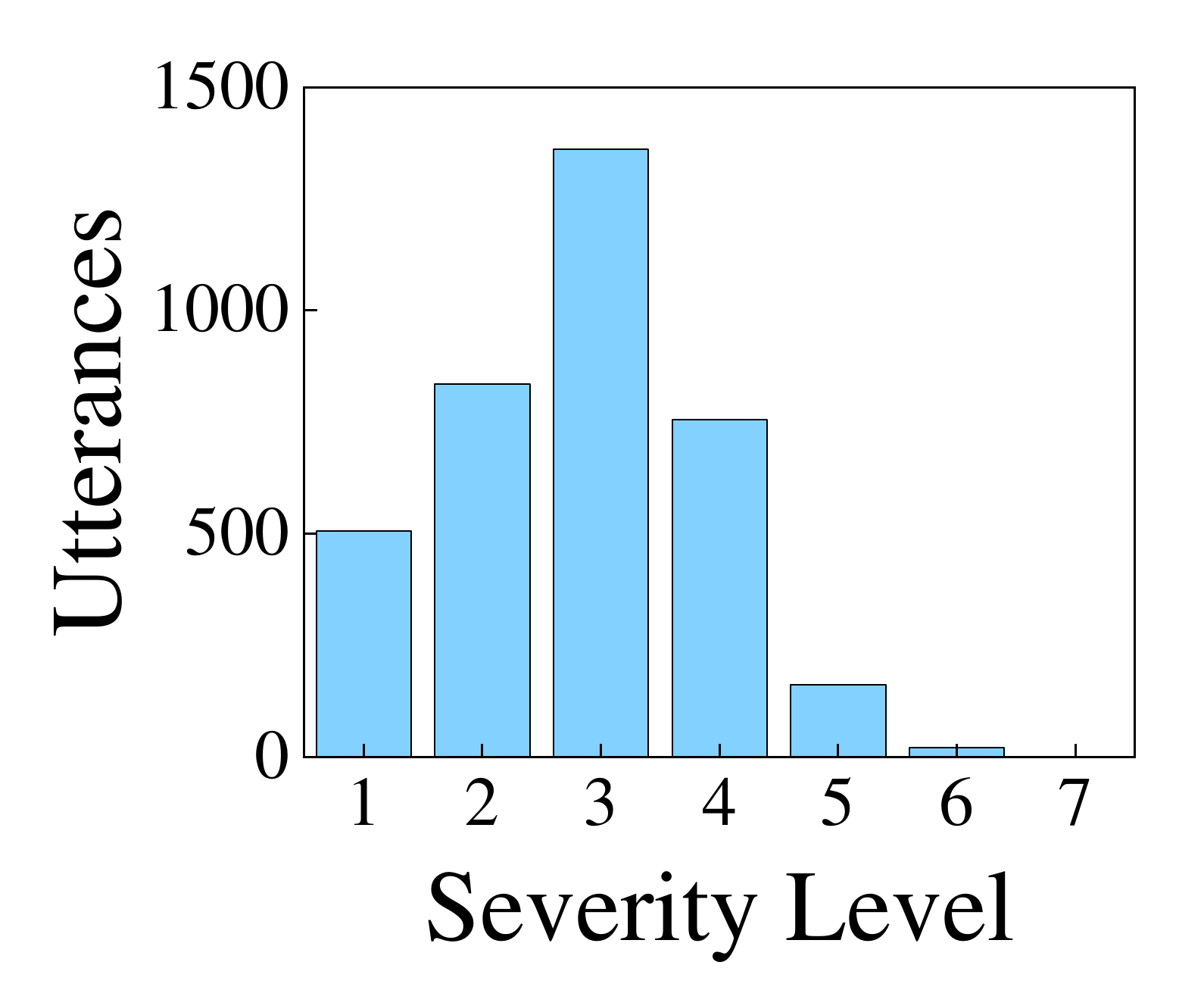}}
    \hfill
    \subfloat[Intelligibility]{\includegraphics[width=0.2\linewidth]{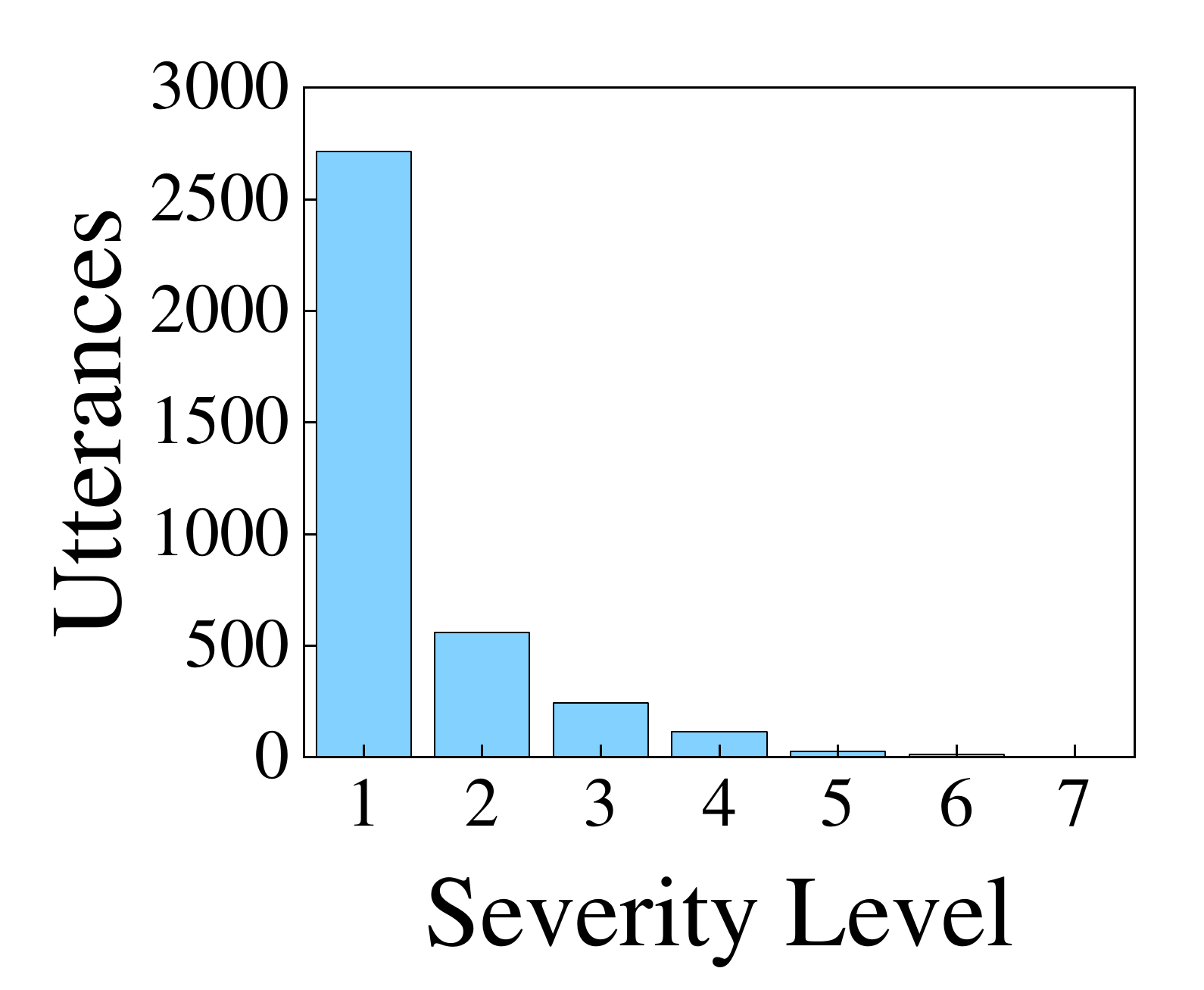}}%
    \hfill
    \subfloat[Imprecise Consonants]{\includegraphics[width=0.2\linewidth]{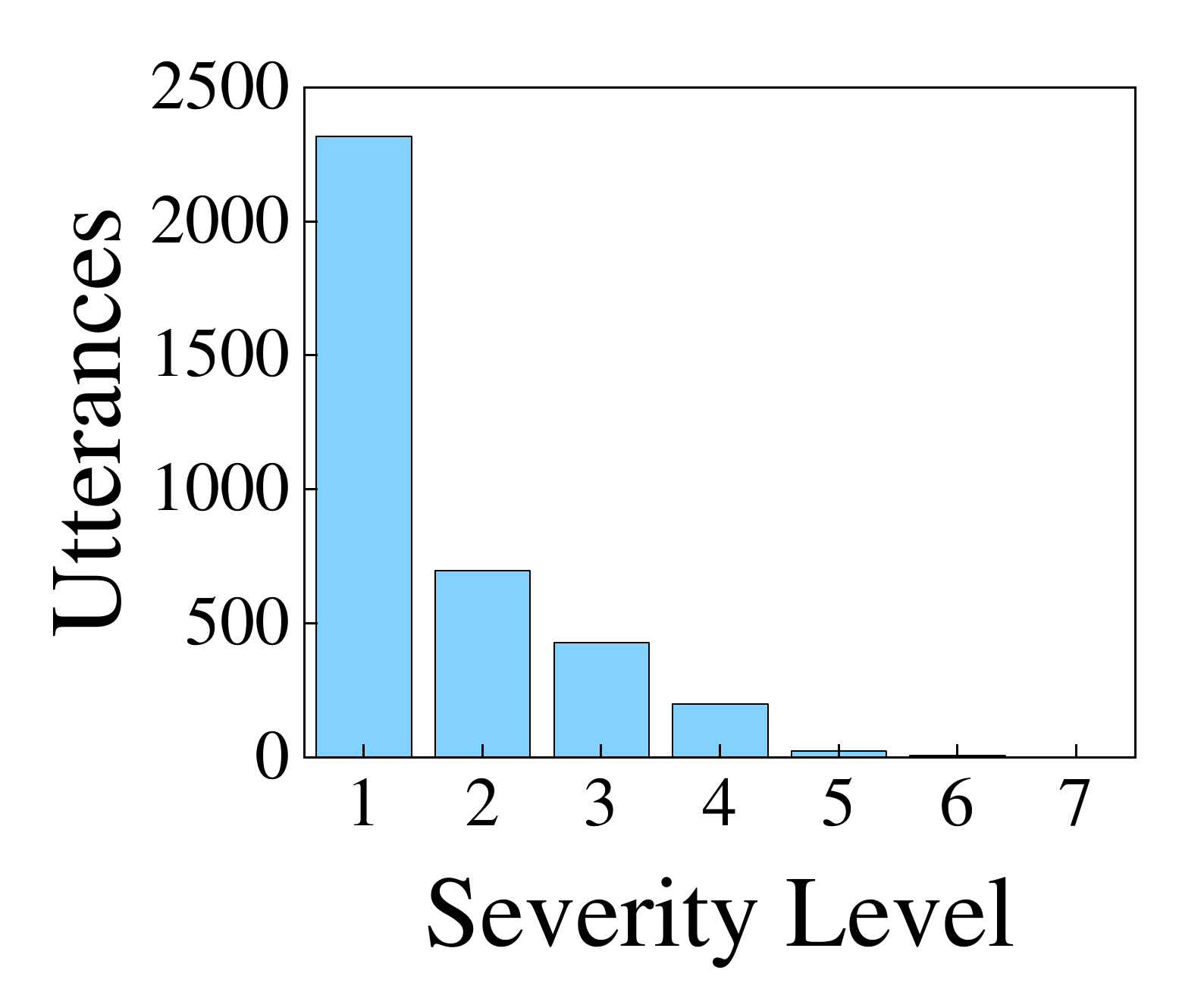}}
    \hfill
    \subfloat[Inappropriate Silences]{\includegraphics[width=0.2\linewidth]{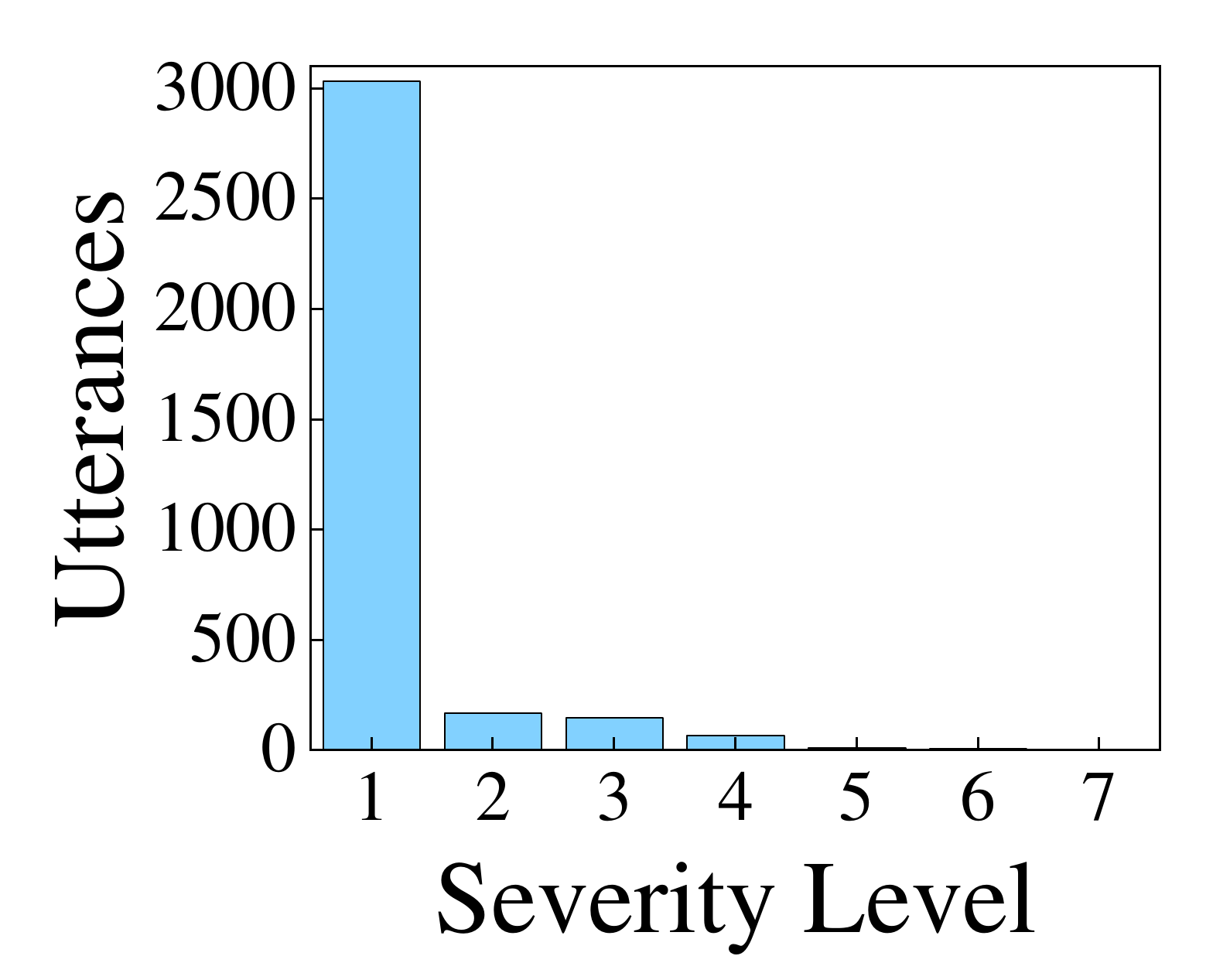}}%
    
    \caption{Distribution of utterances across severity levels (1--7) in the Speech Accessibility Project (SAP) dataset for the five perceptual dimensions: (a) Naturalness, (b) Monopitch, (c) Intelligibility, (d) Imprecise Consonants, and (e) Inappropriate Silences. Higher scores indicate more severe impairment. Notably, several pathological attributes (particularly Inappropriate Silences and Intelligibility) exhibit significant class imbalance toward lower severity levels.}
    \label{fig:layout-221}
\end{figure*}


The rest of this paper is organized as follows.
Section~\ref{sec:task} introduces the utilized corpora and presents task definition.
Section~\ref{sec:method} describes the experiment protocol.
Section~\ref{sec:exp} reports the results and analysis.
Section~\ref{sec:conclusion} concludes the paper.

\section{Task Description}
\label{sec:task}

\subsection{The Speech Accessibility Project Challenge 2025}

The Speech Accessibility Project (SAP) \cite{hasegawa_johnson2024_jslhr} aims to systematically collect, transcribe and distribute speech data from individuals with speech disorders. 
Recorded by participants using their in-home devices, the acoustic environment and task setting of SAP is closer to real-world clinical scenarios.

This work uses a subset of perceptual assessment annotations in the SAP dataset, covering five dimensions: ``Imprecise consonants'', ``Inappropriate silence'', ``Intelligibility'', ``Monopitch'' and ``Naturalness''.
These categories represent some of the most heavily annotated labels in the corpus, providing a balanced coverage that spans both holistic perceptual qualities (i.e., intelligibility and naturalness) and fine-grained phonetic and prosodic dimensions (i.e., consonant precision, silence, and pitch variation).
The above dimensions are annotated by certified speech-language pathologists, using a 1-to-7 severity scale, where higher scores indicate more severe impairment. 

Due to the unreleased perceptual labels in the original test set, the remaining annotated dataset is reorganized into patient-independent splits. 
The original validation set (61 patients) is adopted as the test set, and the original training set (120 patients) is divided into the training (108 patients) and validation (12 patients) subsets. 
Consequently, each dimension contains roughly 2,200 training, 400 validation, and 1,000 testing utterances.

\subsection{QualiSpeech: A Descriptive Corpus for Speech Quality Assessment}

%
%
QualiSpeech is an English corpus developed for multi-dimensional speech quality assessment, designed to complement and refine the conventional single-score MOS paradigm. 
It encompasses diverse acoustic scenarios spanning synthetic speech, channel-degraded speech and authentic in-the-wild recordings.
Each utterance in QualiSpeech provides complete evaluations across seven perceptual dimensions, covering fine-grained attributes (\emph{Noise}, \emph{Distortion}, \emph{Speed}, \emph{Continuity}, \emph{Listening Effort} and \emph{Naturalness}) as well as a holistic \emph{Overall} quality rating. 
All dimensions are scored on a standard 1-to-5 scale, with higher scores reflecting superior perceptual quality. 
All seven dimensions are utilized as multi-task supervision targets exclusively during the pre-training stage. 
Following the official split (10,558 training, 2,167 validation, and 1,782 test samples), its original data distribution is preserved to impart realistic perceptual-quality priors to the model before clinical adaptation.

\section{Method}
\label{sec:method}

\begin{table*}[!ht]
\centering
\caption{Comparison of Auditory Large Language Models (ALLMs) on speech quality prediction over the SAP test set. Spearman ($\rho$) and Pearson ($r$) correlations and Mean Absolute Error (MAE) are computed between model predictions and human ratings on five dimensions: naturalness, monopitch, intelligibility, imprecise consonants, and inappropriate silences. Higher $\rho$ and $r$ and lower MAE indicate better performance. Best result per dimension and metric is in bold.}
\label{tab:baselines}
 \vspace{5pt}
\small
\setlength{\tabcolsep}{5pt}
\begin{tabular}{lcccccccccccc}
\toprule
\multirow{2}{*}{Dimension} & \multicolumn{3}{c}{Qwen2-Audio} & \multicolumn{3}{c}{Qwen2.5-Omni} & \multicolumn{3}{c}{video-SALMONN 2 plus} & \multicolumn{3}{c}{SALMONN 2} \\
\cmidrule(lr){2-4}\cmidrule(lr){5-7}\cmidrule(lr){8-10}\cmidrule(lr){11-13}
 & $\rho\uparrow$ & $r\uparrow$ & MAE$\downarrow$ & $\rho\uparrow$ & $r\uparrow$ & MAE$\downarrow$ & $\rho\uparrow$ & $r\uparrow$ & MAE$\downarrow$ & $\rho\uparrow$ & $r\uparrow$ & MAE$\downarrow$ \\
\midrule
Naturalness             & 0.655 & 0.684 & 0.764 & \textbf{0.828} & \textbf{0.836} & \textbf{0.439} & 0.702 & 0.738 & 0.703 & 0.584 & 0.644 & 0.870 \\
Monopitch               & 0.545 & 0.534 & \textbf{0.627} & \textbf{0.584} & \textbf{0.595} & 0.643 & 0.572 & 0.567 & 0.658 & 0.386 & 0.406 & 0.732 \\
Intelligibility         & 0.549 & 0.548 & 0.366 & 0.610 & 0.596 & 0.350 & \textbf{0.635} & \textbf{0.691} & \textbf{0.322} & 0.592 & 0.661 & 0.338 \\
Imprecise Consonants    & 0.632 & 0.691 & 0.449 & 0.700 & 0.673 & 0.411 & \textbf{0.738} & \textbf{0.789} & \textbf{0.342} & 0.640 & 0.679 & 0.456 \\
Inappropriate Silences  & 0.335 & 0.371 & 0.193 & \textbf{0.631} & \textbf{0.750} & 0.142 & 0.555 & 0.616 & \textbf{0.099} & 0.226 & 0.298 & 0.136 \\
\bottomrule
\end{tabular}
\end{table*}

\begin{table*}[!ht]
\centering
\caption{Effect of post-training on Qwen2.5-Omni for speech quality prediction over the SAP test set: zero-shot performance (Zero-shot), fine-tuned with QualiSpeech (Post-QualiSpeech) and post-training with SAP (Post-SAP).}
\label{tab:posttraining}
\vspace{5pt}
\small
\setlength{\tabcolsep}{5pt}
\begin{tabular}{lccccccccc}
\toprule
\multirow{2}{*}{Dimension} & \multicolumn{3}{c}{Zero-shot} & \multicolumn{3}{c}{Post-QualiSpeech} & \multicolumn{3}{c}{Post-SAP} \\
\cmidrule(lr){2-4}\cmidrule(lr){5-7}\cmidrule(lr){8-10}
 & $\rho\uparrow$ & $r\uparrow$ & MAE$\downarrow$ & $\rho\uparrow$ & $r\uparrow$ & MAE$\downarrow$ & $\rho\uparrow$ & $r\uparrow$ & MAE$\downarrow$ \\
\midrule
Naturalness             & 0.307 & 0.327 & 2.292 & 0.444 & 0.437 & 1.761 & 0.828 & 0.836 & 0.439 \\
Monopitch               & 0.264 & 0.267 & 1.490 & 0.189 & 0.190 & 1.857 & 0.584 & 0.595 & 0.643 \\
Intelligibility         & 0.328 & 0.363 & 1.934 & 0.150 & 0.100 & 2.789 & 0.610 & 0.596 & 0.350 \\
Imprecise Consonants    & 0.492 & 0.547 & 1.714 & 0.419 & 0.463 & 2.215 & 0.700 & 0.673 & 0.411 \\
Inappropriate Silences  & $-$0.073 & $-$0.034 & 3.303 & 0.098 & 0.093 & 3.275 & 0.631 & 0.750 & 0.142 \\
\bottomrule
\end{tabular}
\end{table*}

\subsection{Training Method}

Training proceeds in two stages. The first stage is pre-training on QualiSpeech. The training set of QualiSpeech, containing annotations across all seven dimensions, is mixed, with each sample targeting only one specific dimension. The model undergoes instruction tuning under a causal language modeling objective, learning to map audio perceptual features to discrete MOS scores on the 1-to-5 scale. The second stage is fine-tuning on SAP. The optimal checkpoint selected from Stage 1 is further trained on the dysarthric speech data. For the five target dimensions and the mixed setting, a total of six experimental configurations are conducted. Five are single-dimension experiments, where Imprecise consonants, Inappropriate silence, Intelligibility, Monopitch, and Naturalness are trained independently. The remaining one is the Mix experiment, where training samples from all five dimensions are pooled into a single training set, while each sample is still annotated for only one dimension and the model distinguishes the current evaluation target through dimension-specific prompts. All experiments use the original data distribution without resampling.

\subsection{Evaluation Metrics}

We evaluate the SAP test set at three training checkpoints. Zero-shot uses the original open-source checkpoint directly, serving as a baseline reference. Post-QualiSpeech is evaluated after completing QualiSpeech pre-training, to measure cross-domain transferability from synthetic speech quality assessment to dysarthric speech assessment. Post-SAP is evaluated after completing SAP fine-tuning, reporting the final performance on the SAP test set.

Model performance is assessed at the utterance level using three complementary metrics. Spearman's Rank Correlation Coefficient ($\rho$) measures the rank-order consistency between predicted scores and clinical annotations. Pearson's Linear Correlation Coefficient ($r$) measures the linear correlation strength between predicted and ground-truth scores. Mean Absolute Error (MAE) measures the average absolute deviation between predicted and ground-truth values. During training, Spearman's $\rho$ is used as the checkpoint selection criterion for both Post-QualiSpeech and Post-SAP stages.
\section{Experiments and Results}
\label{sec:exp}

\subsection{Experimental Setup}

We benchmark four representative open-source audio-language models: 
Qwen2-Audio\footnote{\url{https://huggingface.co/Qwen/Qwen2-Audio-7B-Instruct}} \cite{chu2024qwen2audio}, 
Qwen2.5-Omni\footnote{\url{https://huggingface.co/Qwen/Qwen2.5-Omni-7B}} \cite{xu2025qwen25omnitechnicalreport}, 
SALMONN 2\footnote{\url{https://huggingface.co/marcoyang/SALMONN-2-8B}} \cite{yang2026salmonn} 
and video-SALMONN 2 Plus\footnote{\url{https://huggingface.co/tsinghua-ee/video-SALMONN2_plus_3B_full}} \cite{tang2025video}. 
All experiments follow the SAP partition and task formulation defined in Section~\ref{sec:task}. 
Model adaptation uses LoRA with rank $r=8$, scaling factor $\alpha=8$ and dropout of $0.05$. 
Experiments run for 3 epochs with an initial learning rate of $1\times 10^{-5}$, weight decay of $0.01$, and a cosine scheduler with a $5\%$ warmup ratio. 
We train on 4 GPUs with per-device batch size 1 and 4 gradient accumulation steps (effective batch size 16) in \texttt{bfloat16} precision. 
Inference employs greedy decoding. 
Checkpoint selection optimizes Spearman's $\rho$ on the validation set.




\subsection{Multidimensional Assessment Benchmarking}

Table~\ref{tab:baselines} presents the multidimensional assessment performance across all evaluated Auditory Large Language Models under the unified, prompt-conditioned framework (Mix). 
Overall, Qwen2.5-Omni and video-SALMONN 2 Plus establish the strongest baselines, demonstrating compelling complementary advancements across dimensions.

Qwen2.5-Omni excels on both holistic and temporal-prosodic attributes. 
It attains the highest rank and linear correlations and lowest error on \emph{Naturalness} ($\rho=0.828, r=0.836, \text{MAE}=0.439$), alongside the strongest agreement on \emph{Monopitch} ($\rho=0.584, r=0.595$) and \emph{Inappropriate Silences} ($\rho=0.631, r=0.750$). 
Conversely, video-SALMONN 2 Plus demonstrates clear superiority on articulatory precision and communicative clarity, outperforming all counterparts on \emph{Intelligibility} ($\rho=0.635, r=0.691, \text{MAE}=0.322$) and \emph{Imprecise Consonants} ($\rho=0.738, r=0.789, \text{MAE}=0.342$), while maintaining the lowest absolute error on \emph{Inappropriate Silences} ($\text{MAE}=0.099$).

\textbf{Cross-Dimension Error Patterns:}
Performance variations reflect clinical score distributions (Fig.~\ref{fig:layout-221}). 
\emph{Naturalness} exhibits a balanced severity spread, facilitating smooth linear tracking across the continuum. In contrast, attributes reflecting severe neuromotor breakdown (\emph{Intelligibility}, \emph{Imprecise Consonants}, and particularly \emph{Inappropriate Silences}) are heavily concentrated around the unimpaired condition. 
While this sparsity yields deceptively low MAE values (e.g., $0.099$--$0.142$ on silences), the robust rank ($\rho$) and linear ($r$) correlations confirm that leading ALLMs capture fine-grained pathological variations beyond majority-class prediction.

%
%

\begin{figure*}[ht] 
    \centering
    
    \subfloat[]{\includegraphics[width=0.19\linewidth]{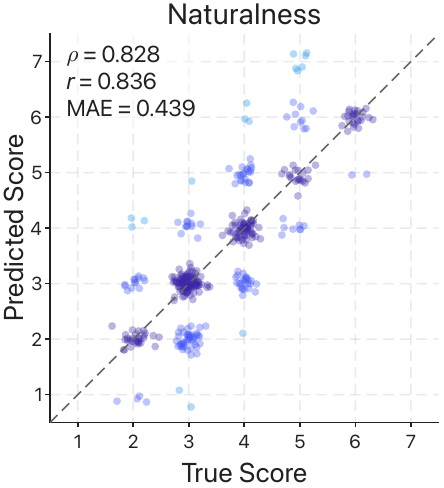}}    
    \hfill 
    \subfloat[]{\includegraphics[width=0.19\linewidth]{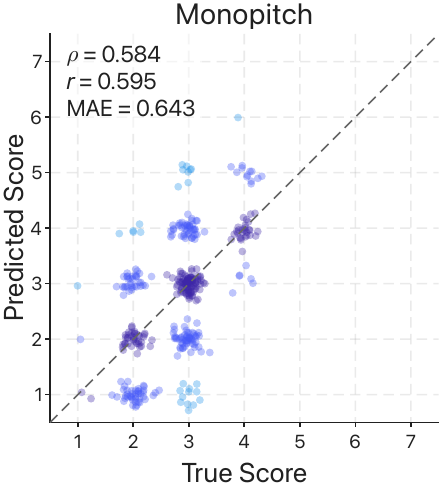}}  
    \hfill
    \subfloat[]{\includegraphics[width=0.19\linewidth]{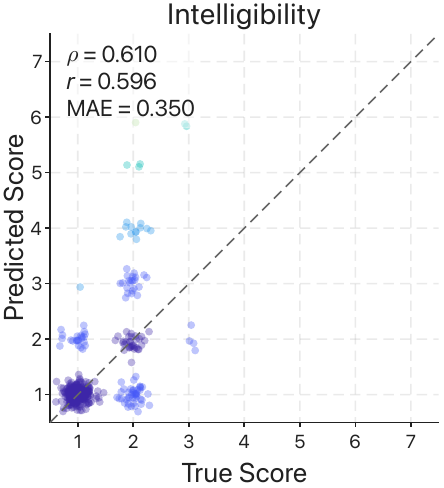}}  
    \hfill
    \subfloat[]{\includegraphics[width=0.19\linewidth]{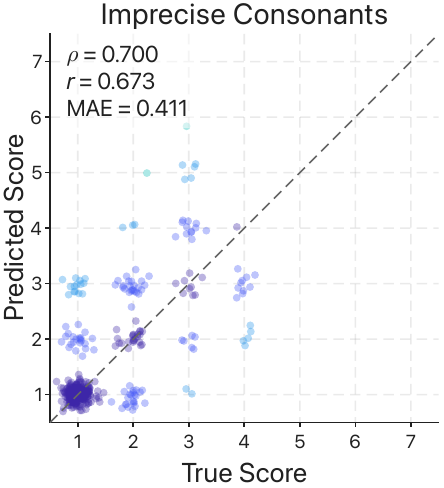}}  
    \hfill
    \subfloat[]{\includegraphics[width=0.19\linewidth]{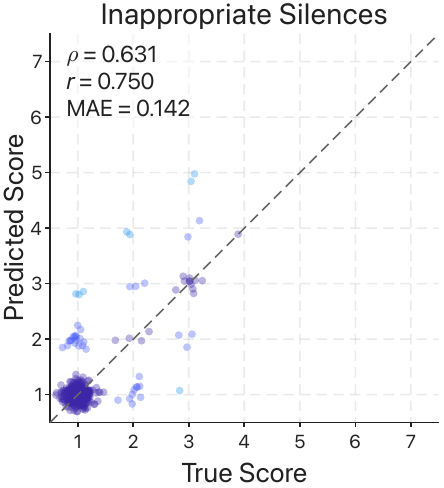}}  
    \caption{Scatter plots of model predictions against human ratings on the SAP test set for (a) Naturalness, (b) Monopitch, (c) Intelligibility, (d) Imprecise Consonants, (e) Inappropriate Silences. These plots visualize the utterance-level predictions corresponding to the Qwen2.5-Omni results reported in Table~\ref{tab:baselines}. Each point represents one test utterance, and the dashed diagonal line indicates perfect agreement between predicted and ground-truth scores.}
    \label{fig:five_images}
\end{figure*}

\begin{table*}[!ht]
\centering
\caption{Systematic comparison of dedicated single-task training (Single) versus unified multi-task training (Mix) across representative Auditory Large Language Models over the SAP test set. video-SALMONN 2 Plus demonstrates substantial multi-task synergy where joint training improves performance across almost all dimensions. Qwen2.5-Omni shows that unified multi-task training achieves competitive accuracy (average $\rho$ within 0.046) with only 1/5 of the parameters, while outperforming single-task models on sparse attributes (Inappropriate Silences) and calibration error (Naturalness MAE). Bold indicates the superior result between Single and Mix for each model.}
\label{tab:ablation}
\vspace{5pt}
\small
\setlength{\tabcolsep}{5pt}
\begin{tabular}{lcccccccccccc}
\toprule
\multirow{2}{*}{Dimension} & \multicolumn{6}{c}{video-SALMONN 2 Plus} & \multicolumn{6}{c}{Qwen2.5-Omni} \\
\cmidrule(lr){2-7}\cmidrule(lr){8-13}
 & \multicolumn{3}{c}{Single} & \multicolumn{3}{c}{Mix} & \multicolumn{3}{c}{Single} & \multicolumn{3}{c}{Mix} \\
\cmidrule(lr){2-4}\cmidrule(lr){5-7}\cmidrule(lr){8-10}\cmidrule(lr){11-13}
 & $\rho\uparrow$ & $r\uparrow$ & MAE$\downarrow$ & $\rho\uparrow$ & $r\uparrow$ & MAE$\downarrow$ & $\rho\uparrow$ & $r\uparrow$ & MAE$\downarrow$ & $\rho\uparrow$ & $r\uparrow$ & MAE$\downarrow$ \\
\midrule
Naturalness             & \textbf{0.722} & \textbf{0.758} & 0.710          & 0.702          & 0.738          & \textbf{0.703} & \textbf{0.848} & \textbf{0.858} & 0.462          & 0.828          & 0.836          & \textbf{0.439} \\
Monopitch               & 0.510          & 0.507          & \textbf{0.633} & \textbf{0.572} & \textbf{0.567} & 0.658          & \textbf{0.693} & \textbf{0.687} & \textbf{0.566} & 0.584          & 0.595          & 0.643          \\
Intelligibility         & 0.592          & 0.638          & 0.346          & \textbf{0.635} & \textbf{0.691} & \textbf{0.322} & \textbf{0.669} & \textbf{0.697} & \textbf{0.311} & 0.610          & 0.596          & 0.350          \\
Imprecise Consonants    & 0.702          & 0.758          & 0.414          & \textbf{0.738} & \textbf{0.789} & \textbf{0.342} & \textbf{0.738} & \textbf{0.736} & \textbf{0.385} & 0.700          & 0.673          & 0.411          \\
Inappropriate Silences  & 0.410          & 0.529          & 0.102          & \textbf{0.555} & \textbf{0.616} & \textbf{0.099} & 0.630          & 0.716          & 0.150          & \textbf{0.631} & \textbf{0.750} & \textbf{0.142} \\
\midrule
Average                 & 0.587          & 0.638          & 0.441          & \textbf{0.640} & \textbf{0.680} & \textbf{0.425} & \textbf{0.716} & \textbf{0.739} & \textbf{0.375} & 0.670          & 0.690          & 0.397          \\
\bottomrule
\end{tabular}
\end{table*}

\subsection{Unified Multi-Task vs. Single-Task Specialization}

A central practical question is whether a single, instruction-conditioned model (\texttt{Mix}) can replace multiple dedicated single-task checkpoints (\texttt{Single}) without suffering severe task interference. 
Table~\ref{tab:ablation} highlights this trade-off across representative architectures.

For video-SALMONN 2 Plus, multi-task joint training yields unambiguous synergistic gains across nearly all targets. 
Transitioning from \texttt{Single} to \texttt{Mix} improves correlation on \emph{Monopitch} ($\rho$ from $0.510$ to $0.572$), \emph{Intelligibility} ($\rho$ from $0.592$ to $0.635$), \emph{Imprecise Consonants} ($\rho$ from $0.702$ to $0.738$), and \emph{Inappropriate Silences} ($\rho$ from $0.410$ to $0.555$). 
On average, \texttt{Mix} elevates $\rho$ by $+0.053$, $r$ by $+0.042$, and reduces MAE by $-0.016$, proving that instruction conditioning successfully captures mutually reinforcing acoustic-phonetic cues.

For Qwen2.5-Omni, specialized single-task models maintain a marginal edge on localized acoustic attributes (\emph{Monopitch} and \emph{Intelligibility}), reflecting minor parameter capacity competition under low-rank LoRA constraints. 
Nevertheless, the unified model achieves highly competitive overall fidelity (mean $\rho=0.670$ vs. $0.716$), while slashing deployment and storage costs by $80\%$ ($1/5$ parameters). 
\texttt{Mix} surpasses dedicated models on the most clinically biased attribute, \emph{Inappropriate Silences} ($\rho=0.631$ vs. $0.630, r=0.750$ vs. $0.716, \text{MAE}=0.142$ vs. $0.150$), and yields lower MAE on \emph{Naturalness} ($\text{MAE}=0.439$ vs. $0.462$). 
Unified prompt-conditioned modeling thus demonstrates strong clinical utility and architectural scalability.

%
%
%
%
\subsection{Cross-Domain Transfer Dynamics from MOS to Pathology}

Table~\ref{tab:posttraining} tracks the progressive capability evolution of Qwen2.5-Omni across training stages to evaluate cross-domain transfer from general perceptual MOS to clinical dysarthria.
Zero-shot foundation models demonstrate rudimentary yet non-negligible auditory sensitivity (e.g., $\rho=0.492$ on consonants), but fail entirely on temporal pathology ($\rho=-0.073$ on silences). 
Pre-training on QualiSpeech induces strongly attribute-dependent transfer: \emph{Naturalness} improves noticeably ($\rho$ from $0.307$ to $0.444$, MAE from $2.292$ to $1.761$), aligning with generic perceptual quality constructs. 
However, performance degrades on motor-speech attributes (\emph{Intelligibility} $\rho$ drops from $0.328$ to $0.150$; \emph{Monopitch} and consonants decline). 
Subsequent in-domain SAP adaptation resolves this disparity, producing noticeable improvements across all dimensions. 
Collectively, these findings confirm that out-of-domain perceptual supervision provides beneficial auditory information for downstream clinical adaptation.

\section{Conclusion}
\label{sec:conclusion}

This paper presents a systematic investigation into auditory large language models (ALLMs) for multi-dimensional perceptual assessment of dysarthric speech. 
Extensive benchmarking across diverse architectures demonstrates that leading ALLMs can reliably quantify both holistic and fine-grained pathological dimensions, showing complementary strengths across prosodic and articulatory attributes. 
Furthermore, a unified, prompt-conditioned multi-task formulation achieves competitive or superior fidelity compared to dedicated single-task models while drastically reducing deployment overhead. 
Additionally, cross-domain perceptual pre-training offers beneficial auditory priors that facilitate subsequent in-domain clinical adaptation. 
Overall, the findings validate the strong potential of ALLMs as unified clinical evaluators and underscore that effective dysarthric speech assessment requires jointly aligning model selection, perceptual adaptation, and task formulation.

\vfill\pagebreak

\bibliographystyle{IEEEbib}
\bibliography{refs}

\end{document}